\documentstyle[12pt]{article}
\title{\bf 't Hooft Loop Average\\ 
in the Vicinity of the Londons' Limit and the\\
Quartic Cumulant of the Field Strength Tensors}
\author{D.V.ANTONOV \thanks{ E-mail addresses:
antonov@pha2.physik.hu-berlin.de, antonov@vxitep.itep.ru}{\,}
\thanks {Supported
by Graduiertenkolleg {\it Elementarteilchenphysik}, Russian
Fundamental Research Foundation, Grant No.96-02-19184, DFG-RFFI,
Grant 436 RUS 113/309/0, and by the INTAS, Grant No.94-2851.}
\\
{\it Institute of Theoretical and Experimental Physics,}\\
{\it B.Cheremushkinskaya 25, 117 218, Moscow, Russia}\\
{\it and}\\
{\it Institut f\"ur Physik, Humboldt-Universit\"at,}\\
{\it Invalidenstrasse 110, D-10115, Berlin, Germany}}
\date{}
\begin{document}
\maketitle
\vspace{1mm}
\centerline{\bf {Abstract}}
\vspace{3mm}

The next-to-leading term in the weight factor of the string representation 
of the 't Hooft 
loop average defined on the string world-sheet is found in the Abelian 
Higgs Model near the Londons' limit. 
This term emerges due to the finiteness of the coupling constant and,  
in contrast to the Londons' limit, where only the bilocal 
cumulant in the expansion of the 't Hooft average survived, 
leads to the appearance of the quartic 
cumulant. Apart from 
the Londons' penetration depth of the 
vacuum, which was a typical fall-off scale of the bilocal cumulant, 
the quartic cumulant  
depends also on the other characteristic length of the Abelian Higgs 
Model, the correlation radius of the Higgs field.

\newpage

In a recent paper$^{1}$, a string representation for the 't Hooft 
loop average defined on the string world-sheet 
in the Londons' limit of the Abelian Higgs Model (AHM) 
has been constructed. This has been done by virtue of the duality 
transformation, proposed in Refs. 2 and 3. 
It has been also shown that in this limit, 
the bilocal approximation to the Method of Vacuum Correlators$^{4}$ 
is an exact result, so that all the cumulants higher than quadratic 
one vanish. However, while in Ref. 1 it has been demonstrated that the 
Londons' limit is consistent according to the lattice data for the 
string tension and the correlation length of the vacuum in QCD$^{5}$, 
it has been argued in Ref. 4 that in the real QCD vacuum all 
the higher cumulants are 
nonzero as well (though the bilocal cumulant is in fact the dominant one). 
Therefore it looks natural to obtain a correction to the weight 
factor of the string 
representation of the 't Hooft loop average, found in Ref. 1, 
emerging due to 
the finiteness of the coupling constant, which must yield the next
term in the cumulant expansion. This is just the problem which 
will be studied in the present Letter. It is also worth mentioning,  
that the finiteness of the coupling constant has already been 
taken into account during the derivation of the monopole effective action 
in the dual Abelian Higgs Model in Ref. 6.      

In order to get a desirable correction to the weight factor 
of the string representation of the 't Hooft loop average, 
let us expand the radial part of the 
Higgs wave function, $\Phi(x)=\varphi(x){\rm e}^{i\theta(x)}$, as 
$\varphi(x)=\eta+\tau\psi(x)$, where $\theta=\theta^{{\rm sing.}}+
\theta^{{\rm reg.}}$, $\eta$ stands for the square 
root of the Higgs field vacuum expectation value (from now on we shall 
make use of notations of Refs. 1 and 3),  $\tau\equiv\frac{1}{\lambda}
\to 0$, and $\psi(x)$ is an arbitrary quantum fluctuation. 
Obviously, in what 
follows we shall keep only the terms linear in $\tau$. Then the weight 
factor in the string representation of the 't~Hooft 
loop average has the form 

$$\left<{\cal F}_N(S)\right>\equiv 
\int \varphi{\cal D}\varphi{\cal D}A_\mu
{\cal D}\theta^{{\rm reg.}}\exp\Biggl\{-\int dx\Biggl[\frac14 F_{\mu\nu}^2+
\frac{\tau^2}{2}\left(\partial_\mu\psi\right)^2+4\tau\eta^2
\psi^2+$$

$$+\eta\left(\frac{\eta}{2}+\tau\psi\right)\left(\partial_\mu
\theta-NeA_\mu\right)^2+\frac{\pi}{Ne}\varepsilon_{\mu\nu\alpha\beta}
T_{\mu\nu}F_{\alpha\beta}\Biggr]\Biggr\}, $$
where 

$$T_{\mu\nu}(x)\equiv\int\limits_S^{} 
d\sigma_{\mu\nu}(x(\xi))\delta(x-x(\xi))=\frac{1}{2\pi}
\varepsilon_{\mu\nu\rho\sigma}\partial_\rho\partial_\sigma
\theta^{{\rm sing.}}(x)$$ 
is the vorticity tensor current (the last equality is due to Ref. 2), and  
$S$ is a closed string world-sheet 
parametrized by $x_\mu(\xi)$. 
The integral $\int {\cal D}x_\mu(\xi)
\left<{\cal F}_N(S)\right>$ is a constant, which  
by definition is called  
the 't Hooft loop average defined 
on the string world sheet, but in what follows, we shall be 
interested in the 
weight factor  
$\left<{\cal F}_N(S)\right>$ rather than in this constant. 

One's first instinct is to claim that this weight factor is independent 
of the shape of $S$, but depends only on the hypersurface ${\cal V}$ bounded 
by $S$, since one may think that due to the Gauss theorem, 
$\left<{\cal F}_N(S)\right>$ is equal to

$$\left<\exp\left(\frac{4\pi}{Ne}\int\limits_{{\cal V}}^{} dV_\mu 
\partial_\nu F_{\nu\mu}\right)\right>.$$
However, this is wrong, since the $S$-dependence is present in 
$\left<{\cal F}_N(S)\right>$ not only in the last term, but also in the 
term $\partial_\mu
\theta^{{\rm sing.}}$, and the naive application of the 
Gauss theorem is therefore not valid. 

Let us stress, that while the usual 't Hooft loop average defined on 
a {\it fixed closed} surface is just some constant, this is 
not the case for the weight factor $\left<{\cal F}_N(S)\right>$, 
since $\left<...\right>$ here means only the average 
over all the fields except $\theta^{{\rm sing.}}$ (for simplicity, we 
do not take into account the Jacobian arising when one passes from the 
integration over $\theta^{{\rm sing.}}$ to the integration over $x_\mu
(\xi)$). 
As it was already mentioned above, 
in what follows, our main goal will be the search 
for the irreducible correlators (cumulants) of the dual field strength 
tensors. To this end, we shall compare   
$\left<{\cal F}_N(S)\right>$ expressed via $x_\mu(\xi)$  
with the cumulant expansion of this object. Notice that, if  
we would, instead of that, add external 
monopoles into the theory and consider the usual 't Hooft loop average 
defined on an arbitrary open surface encircled by the monopole ring
(which is not a constant any more, but a functional of this open surface), we 
would get in the weight factor of its string representation  
correlations between surface elements lieing both on this surface and 
on the string world-sheet. This would make it difficult to compare 
this weight factor with its cumulant expansion. 
Therefore, we do not consider any external monopoles, and are 
consequently left with the correlations between the elements of the 
unique surface in the system under study, $S$.

Performing the same duality transformation as the one which has been 
proposed in Ref. 2 and applied in Refs. 1 and earlier in 3, we arrive at the  
following expression for $\left<{\cal F}_N(S)\right>$  

$$\left<{\cal F}_N(S)\right>=
\int {\cal D}h_{\mu\nu}
\exp\Biggl\{\int dx\Biggl[-\frac{1}{12\eta^2}H_{\mu\nu\lambda}^2-
\left(\frac{Ne}{2}\right)^2h_{\mu\nu}^2+\left(\frac{2\pi}{Ne}\right)^2
T_{\mu\nu}^2+3\pi i h_{\mu\nu}T_{\mu\nu}\Biggr]\Biggr\}\cdot$$

$$\cdot\int\varphi {\cal D}\varphi\exp\Biggl\{\int dx\Biggl[
-\frac{\tau^2}{2}\left(\partial_\mu\psi\right)^2-4\tau\eta^2
\psi^2+\frac{\tau\psi}{6\eta^3}H_{\mu\nu\lambda}^2
\Biggr]\Biggr\}, \eqno (1)$$
where $H_{\mu\nu\lambda}=\partial_\mu h_{\nu\lambda}+\partial_\lambda 
h_{\mu\nu}+\partial_\nu h_{\lambda\mu}$ 
is a stength tensor of the massive Kalb-Ramond 
field $h_{\mu\nu}$. The exponent standing in the first line 
of Eq. (1) corresponds to the Londons' 
limit (it coincides with the one in Eq. (6) of Ref. 1), while the 
exponent standing in the second line is the desirable 
correction emerging due to the finiteness of 
$\lambda$.

Neglecting the Jacobian, which arises when one passes from the 
integration over $\varphi$ to the integration over $\psi$, 
and keeping in the expansion of $\varphi$ standing in 
the preexponent only the first term, $\eta$, it is easy to carry 
out the Gaussian integral, which stands in the second line of Eq. (1). 
The result has the form 

$$\exp\Biggl[\frac{\sqrt{\frac{2}{\tau}}}{144\pi^2\eta^5}\int dxdy
\frac{K_1\left(2\sqrt{\frac{2}{\tau}}\eta\left|x-y\right|\right)}
{\left|x-y\right|}H_{\mu\nu\lambda}^2(x)H_{\alpha\beta\gamma}^2(y)
\Biggr], \eqno (2)$$
where from now on $K_i$'s, $i=0,1,2$, are the Macdonald functions, 
and it is worth noting that the characteristic distance 
$\frac{1}{2\sqrt{2\lambda}\eta}$, at which the integrand in 
Eq. (2) vanishes, is exactly the correlation radius of the 
Higgs field, 
$\frac{1}{m_{\rm Higgs}}$.   
Notice also, that one can see that the $\frac{1}{\lambda}$-expansion 
leads already in 
its lowest order to the nonlocal action.

In order to get the contribution of the term (2)  
to the cumulant expansion of 
$\left<{\cal F}_N(S)\right>$, let us 
substitute into Eq. (2) the saddle-point value of the field $h_{\mu\nu}$, 
following from the exponent standing in the first line of 
Eq. (1), i.e. the value of this field corresponding  
to the Londons' limit, which reads 

$$h_{\mu\nu}^{\rm extr.}(x)=
\frac{3i\eta}{2\pi}\Biggl\{m\eta\int\limits_S^{} d\sigma_{\mu\nu}
(y)\frac{K_1}{\left|x-y\right|}+$$

$$+\frac{1}{Ne}\Biggl[{\,}\oint\limits_{\partial S}^{} dy_\mu
(x-y)_\nu-\oint\limits_{\partial S}^{} 
dy_\nu (x-y)_\mu\Biggr]\frac{1}{(x-y)^2}\Biggl[
\frac{K_1}{\left|x-y\right|}+\frac{m}{2}\Biggl(K_0+K_2\Biggr)\Biggr]
\Biggr\}, \eqno (3)$$
where $m\equiv Ne\eta$ is the mass of the Kalb-Ramond field, 
equal to the mass of the gauge boson generated by the Higgs 
mechanism, and 
the arguments of all the Macdonald functions are the same, $m\left|x-
y\right|$. Since the surface $S$ is closed, the boundary terms,  
standing in the last line of Eq. (3), as well as 
the contribution of the function $D_1\left(m^2x^2\right)$, given by 
Eq. (12) of Ref. 1, to the string effective action vanish.  
This remark concerning the function $D_1$ 
means that the formal manipulations with 
AHM in the Londons' limit allow one to obtain both functions $D$ and 
$D_1$ which parametrize the bilocal cumulant, 
but however actually the contribution 
of only one of them 
survives after the integration over the closed surface.  

Substituting Eq. (3) into Eq. (2), we get the following contribution of the 
quartic cumulant of the dual field strength tensors to the
weight factor $\left<{\cal F}_N(S)\right>$

$$\exp\Biggl[\frac{1}{4!}\left(\frac{2\pi}{Ne}\right)^4
\int\limits_S^{} 
d\sigma_{\nu\lambda}(x')d\sigma_{\rho\mu}(x'')d\sigma_{\beta\gamma}
(y')d\sigma_{\sigma\alpha}(y'')\left<\left<\tilde F_{\nu\lambda}(x')
\tilde F_{\rho\mu}(x'')\tilde F_{\beta\gamma}(y')\tilde F_{\sigma
\alpha}(y'')\right>\right>\Biggr],$$ 
where the cumulant itself reads as follows

$$\left<\left<\tilde F_{\nu\lambda}(x')\tilde F_{\rho\mu} (x'')
\tilde F_{\beta\gamma}(y')\tilde F_{\sigma\alpha}(y'')\right>\right>=
\frac{243\sqrt{2}}{512\pi^{10}}\sqrt{\lambda}\left(Ne\right)^8\eta^7
\delta_{\rho\nu}\delta_{\sigma\beta}\cdot$$

$$\cdot\int dxdy\Biggl[4(y'-y)_\alpha
(y''-y)_\gamma\Biggl((x'-x)_\mu(x''-x)_\lambda-\delta_{\mu\lambda}
(x'-x)_\zeta(x''-x)_\zeta\Biggr)+$$

$$+\delta_{\mu\lambda}\delta_{\alpha
\gamma}(x'-x)_\zeta (x''-x)_\zeta (y'-y)_\chi (y''-y)_\chi\Biggr]
\frac{K_1\left(m_{\rm Higgs}\left|x-y\right|\right)}{\left|x-y\right|}
\cdot$$

$$\cdot f\left(\left|x'-x\right|\right)f\left(\left|x''-x\right|\right)
f\left(\left|y'-y\right|\right)f\left(\left|y''-y\right|\right) 
\eqno (5)$$
with 

$$f\left(\left|x\right|\right)\equiv\frac{1}{x^2}\Biggl[\frac{K_1
\left(m\left|x\right|\right)}{\left|x\right|}+\frac{m}{2}\Biggl(
K_0\left(m\left|x\right|\right)+K_2\left(m\left|x\right|\right)
\Biggr)\Biggr]. \eqno (6)$$

Eqs. (5) and (6), which yield the value of the quartic cumulant, 
are the main result of the present Letter. We would like to stress once 
more that besides the gauge boson mass $m$, in Eq. (5)  
there appeared also the other mass parameter of AHM, the Higgs mass 
$m_{\rm Higgs}$. It is also worth mentioning, that the threelocal 
cumulant occured to be equal to zero, which according to Ref. 7 
should presumably imply that all the odd terms in the cumulant 
expansion of $\left<{\cal F}_N(S)\right>$ vanish.

In conclusion, we have proved the conjecture, which has been put 
forward in Ref. 1, 
that the bilocal approximation to the Method of Vacuum Correlators 
is exact only in the Londons' limit of AHM, whereas accounting for 
the finiteness of $\lambda$ leads to the appearance of the higher 
cumulants. It would be interesting to study the contributions of the 
cumulant (5),(6) to the string tension of the Nambu-Goto term and 
to the bare coupling constant of the rigidity rerm, which will be 
the topic of the next publication.   

\vspace{6mm}
{\large \bf Acknowledgments}

\vspace{3mm}

The author is deeply grateful to Professors 
M.I.Polikarpov and Yu.A.Simonov 
for useful discussions, and to M.N.Chernodub for discussions 
and criticism. He would also like 
to thank the theory group of the Quantum Field Theory Department of the 
Institute of Physics of the Humboldt University of Berlin and 
especially Profs. D.Ebert, D.L\"ust, and M.M\"uller-Preussker   
for kind hospitality. 

\vspace{6mm}
{\large \bf References}

\vspace{3mm}
\noindent
1.~D.V.Antonov, {\it hep-th}/9710144.\\
2.~K.Lee, {\it Phys.Rev.} {\bf D48}, 2493 (1993).\\ 
3.~M.I.Polikarpov, U.-J.Wiese, and M.A.Zubkov, {\it Phys.Lett.} 
{\bf B309}, 133 (1993); P.Orland, {\it Nucl.Phys.} {\bf B428}, 
221 (1994), M.Sato and S.Yahikozawa, {\it Nucl.Phys.} {\bf B436}, 
100 (1995); 
E.T.Akhmedov, M.N.Chernodub, M.I.Polikarpov, and M.A.Zubkov, 
{\it Phys.Rev.} {\bf D53}, 2087 (1996); E.T.Akhmedov, {\it JETP Lett.} 
{\bf 64}, 82 (1996).\\
4.~H.G.Dosch, {\it Phys.Lett.} {\bf B190}, 177 (1987); Yu.A.Simonov,
{\it Nucl.Phys.} {\bf B307}, 512 (1988); H.G.Dosch and Yu.A.Simonov,
{\it Phys.Lett.} {\bf B205}, 339 (1988), {\it Z.Phys.} {\bf C45}, 147
(1989); Yu.A.Simonov, {\it Nucl.Phys.} {\bf B324}, 67 (1989), {\it
Phys.Lett.} {\bf B226}, 151 (1989), {\it Phys.Lett.} {\bf B228}, 413
(1989), {\it Yad.Fiz.} {\bf 58}, 113 (1995), preprint ITEP-PH-97-4 
({\it hep-ph}/9704301); for a review see Yu.A.Simonov, 
{\it Yad.Fiz.} {\bf 54}, 192 (1991), {\it Phys.Usp.} 
{\bf 39}, 313 (1996).\\
5.~A. Di Giacomo and H.Panagopoulos, {\it Phys.Lett.} {\bf B285}, 133 
(1992); A. Di Giacomo, E.Meggiolaro, and H.Panagopoulos, 
{\it hep-lat}/9603017 (preprints IFUP-TH 12/96 and UCY-PHY-96/5) 
(in press in {\it Nucl.Phys.} {\bf B}).\\ 
6.~S.Kato, M.N.Chernodub, S.Kitahara, N.Nakamura, M.I.Polikarpov, and 
T.Suzuki, preprint KANAZAWA-97-15 ({\it hep-lat}/9709092); 
M.N.Chernodub and M.I.Polikarpov, preprint ITEP-TH-55-97 
({\it hep-th}/9710205).\\
7.~Yu.A.Simonov and V.I.Shevchenko, {\it Phys.Atom.Nucl.} {\bf 60}, 
1201 (1997).

\end{document}